\documentclass{emulateapj}
\usepackage{graphicx}
\usepackage{graphics}

\slugcomment{}

\shorttitle{}
\shortauthors{} 

\begin{document}

\title{The early blast wave of the 2010 explosion of U Scorpii}

\author{J.J. Drake}
\affil{Harvard-Smithsonian Center for Astrophysics, 60 Garden
              Street, Cambridge, MA 02138, USA}

\author{S. Orlando}
\affil{INAF - Osservatorio Astronomico di Palermo ``G.S.
              Vaiana'', Piazza del Parlamento 1, 90134 Palermo, Italy}

%\email{}

\begin{abstract}
Three-dimensional hydrodynamic simulations exploring the first
18 hours of the 2010 January 28 outburst of the recurrent nova U~Scorpii
have been performed. Special emphasis was placed on capturing the
enormous range in spatial scales in the blast. The pre-explosion system
conditions included the secondary star and a flared accretion disk.
These conditions can have a profound influence on the
evolving blast wave. The blast itself is shadowed by the secondary star,
which itself gives rise to a low-temperature bow-shock. The accretion
disk is completely destroyed in the explosion. A model with a 
disk gas density of $10^{15}$~cm$^{-3}$ produced a blast wave that is
collimated and with clear bipolar structures, including a bipolar
X-ray emitting shell. The degree of collimation depends on the
initial mass of ejecta, energy of explosion, and circumstellar gas density distribution.  It is most pronounced for
a model with the lowest explosion energy ($10^{43}$ erg) and mass of
ejecta ($10^{-8}M_{\odot}$). The X-ray luminosities of three of six models
computed are close to, but consistent with, an upper limit to the early
blast X-ray emission obtained by the {\it Swift} satellite, the
X-ray luminosity being larger for higher circumstellar gas density and 
higher ejecta mass.  The latter consideration, together with estimates of the blast energy from previous outbursts, 
suggests that the mass of ejecta in the 2010 outburst
was not larger than $10^{-7}~M_{\odot}$.
\end{abstract}


\section{Introduction}
\label{s:intro}

U~Scorpii was discovered to have entered its latest outburst on 2010 January 28 by amateur astronomer B.G.~Harris \citep{Schaefer.etal:10}.
It is one of the best observed of the 10 recurrent novae known to date \citep[e.g.][]{Schaefer:10}, with a fairly constant recurrent cycle of 8--12 yr and a previous outburst in 1999.  It comprises a white dwarf of mass $1.55\pm 0.24 M_\odot$ and late-type star of mass $0.88\pm 0.17 M_\odot$  in an orbit with a period of 1.23d \citep{Johnston.Kulkarni:92, Thoroughgood.etal:01, Schaefer:90,  Schaefer.Ringwald:95}.  \citet{Schaefer:10} adopts a median spectral type of the secondary from estimates culled from the literature of G5~IV.

The outburst mechanism for recurrent novae is thermonuclear runaway on the white dwarf triggered by the accreted mass exceeding a critical limit \citep{Starrfield.etal:85, Starrfield.etal:88}.   The relatively short 10 year cycle of U~Sco and the fast evolution of its outbursts implies the white dwarf is close to the Chandrasekhar limit \citep[e.g.][]{Starrfield.etal:88,Hachisu.etal:00}.  Ejecta from previous outbursts were also found to be He-rich (\citealt{Williams.etal:81, Barlow.etal:81, Anupama.Dewangan:00}; though this was contested by \citealt{Iijima:02}), further fueling discussion of U~Sco as a possible progenitor of a Type~Ia supernova (e.g. \citealt{Starrfield.etal:88,Hachisu.etal:99,Schaefer:10}, see, however, the evolutionary considerations of \citealt{Sarna.etal:06}): \citet{Thoroughgood.etal:01} characterised it as ``the best Type Ia supernova progenitor currently known".

%The 2010 outburst of U~Sco triggered an international multi-wavelength campaign

The 2006 explosion of the recurrent nova RS~Oph provided a cogent demonstration of the diagnostic power of  prompt X-ray observations.  RS~Oph was a particularly bright X-ray source: the explosion occurs within the dense wind of the secondary red giant star, resulting in a $\sim 10^7$--$10^8$~K  blast wave whose evolution was followed by RXTE, Swift  and {\it Chandra} \citep{Sokoloski.etal:06,Bode.etal:06,Nelson.etal:08,Drake.etal:09}.   Detailed 3D hydrodynamic modelling of the blast by \citet{Orlando.etal:09} provided estimates of the explosion energy, ejecta mass, and confirmed the presence of enhanced gas density in the equatorial plane of the system.   

Spectroscopy of the 2010 outburst of U~Sco found
H$_\alpha$ and H$_\beta$ H Balmer line profiles broadened by
7600~km~s$^{-1}$ full-width at half-maximum \citep{Anupama:10}, and by
11000~km~s$^{-1}$ full-width at zero intensity \citep{Arai.etal:10},
consistent with those observed in earlier outbursts
\citep[e.g.][]{Barlow.etal:81,Munari.etal:99,Anupama.Dewangan:00,Iijima:02}.
If partially converted to thermal energy, analogous to the explosion of
RS~Oph into the dense wind of the secondary star, such kinetic energy
would result in heating to X-ray emitting temperatures.  While the
secondary of U~Sco lacks the cool, dense wind of the RS~Oph secondary
that provided the medium for copious X-ray production, the possibility
of the blast generating significant X-rays from shocked ejecta or shocked
gas in its accretion disk or ambient circumbinary material remained.

The early outburst was observed by the {\it Swift}, RXTE and {\it Integral} satellites within a day of discovery \citep{Schlegel.etal:10,Manousakis.etal:10} but all failed to detect any X-ray emission. \citet{Schlegel.etal:10} placed the most stringent upper limit 
of $L_X < 1.1\times 10^{-13}$~erg~s$^{-1}$cm$^{-2}$ (90\%\ confidence) in the 0.3-10 keV band based on data obtained on 2010 Jan 29-30.
%Integral 10 mCrab in 3-10 keV energy band 

Here we describe detailed hydrodynamic simulations of the U~Sco explosion similar to those undertaken for RS~Oph by \citet{Orlando.etal:09}.  We investigate the effects of the accretion disk and close secondary companion on the explosion, and examine possible constraints the lack of observed X-rays might provide.   
%We use the results of the simulations to predict the like future evolution of the remnant %and the likelihood of observing collimated or jet-like features.

%Mwd = 1.55 ± 0.24 M ? for the white dwarf and M2 = 0.88 ± 0.17 M ? for a 
%secondary star. 

\section{Hydrodynamic Modelling}
\label{s:obsanal}

The 3-D hydrodynamic model adopted here is similar to that
of \citet{Orlando.etal:09}. The calculations were performed using FLASH, an
adaptive mesh refinement multiphysics code for astrophysical
plasmas \citep{Fryxell.etal:00}. The hydrodynamic equations for
compressible gas dynamics are solved using the FLASH implementation of the
piecewice-parabolic method (PPM; \citealt{Colella.Woodward:84}). Unlike
the models of \citet{Orlando.etal:09} computed for the case of RS~Oph,
for the very early phase of the blast studied here we do not take into
account radiative losses and thermal conduction, assuming that the
remnant is in its adiabatic expansion phase for the whole evolution
considered. This assumption is valid if the time covered by our
simulations $t_{\rm end}$ is smaller than the transition time from
adiabatic to radiative phase for an expanding spherical blast, defined as
(e.g. \citealt{Blondin.etal:98, Petruk:05}),
\begin{equation}
t_{\rm tr} = 0.522\;E^{4/17}\;n_{\rm med}^{-9/17}~\mbox{s}~,
\label{trans_time}
\end{equation}
where $E=E_{\rm 0}$ is the energy of the explosion and $n_{\rm
med}$ is the particle number density of the ambient medium. As discussed
below, it turns out that $t_{\rm end} \ll t_{\rm tr}$ in our set of
simulations and our assumption is valid.

\begin{table*}[htdp]
\caption{Adopted parameters and initial conditions for the hydrodynamic models of the 2010 U~Sco explosion
\label{t:params}}
\begin{center}
\begin{tabular}{lcccccc}
\hline \hline
Parameter & \multicolumn{6}{c}{Value} \\\hline 
Secondary star radius & \multicolumn{6}{c}{$R_b=2.1 R_\odot$} \\
Binary separation & \multicolumn{6}{c}{$a= 6.5 R_\odot$} \\
Inclination & \multicolumn{6}{c}{$i=83\deg$} \\
Model abbreviation & {\footnotesize E44M-7D13} & {\footnotesize E44M-7D15} & {\footnotesize E43M-8D15} &
{\footnotesize E44M-7D15HW} & {\footnotesize E44M-7D15LW} & 
 {\footnotesize E45M-6D15LW} \\
%  & {\footnotesize D13} & {\footnotesize D15} & {\footnotesize D15} &  {\footnotesize D15HW} & {\footnotesize D15LW} & 
% {\footnotesize D15LW} \\
\hline
Explosion energy $E_0$ (erg) & $10^{44}$  & $10^{44}$  &
$10^{43}$ & $10^{44}$  & $10^{44}$ & $10^{45}$ \\
Ejecta mass $M_{ej}$ ($M_\odot$) & $10^{-7}$ & $10^{-7}$ & $10^{-8}$  & $10^{-7}$ &  $10^{-7}$ & $10^{-6}$  \\
Accn.\ disk density $n_d$ (cm$^{-3}$) & $10^{13}$  &
$10^{15}$  & $10^{15}$  & $10^{15}$ & $10^{15}$ & $10^{15}$  \\ 
Sec.\ wind density $n_w$ (cm$^{-3}$) & $10^6$ & $10^6$ &
 $10^6$ & $10^7$  & $10^{5}$ & $10^{5}$ \\
Eq.\ enhancement  $n_{eq}$ (cm$^{-3}$) & $10^8$ & $10^8$ &
 $10^8$ & $10^9$ & $10^{7}$ & $10^{7}$ \\ 
X-ray luminosity $L_X$ (erg~s$^{-1}$) & $1\times 10^{33}$ & $5\times 10^{32}$ & $3\times 10^{31}$ & $4\times 10^{33}$ & $5\times 10^{31}$ & $7\times 10^{33}$ \\ %\hline
Swift XRT $L_X$ (erg~s$^{-1}$) &  \multicolumn{6}{c}{$< 2\times 10^{33}$ (90\%\ confidence)} \\ \hline 
 
 Spatial domain & \multicolumn{4}{c}{$-6.6 \leq x \leq 6.6$ AU} &
  \multicolumn{2}{c}{$-13.2\leq x \leq 13.2$ AU} 
 \\
  & \multicolumn{4}{c}{$0 \leq y \leq 6.6$ AU} &
  \multicolumn{2}{c}{$0 \leq y \leq 13.2$ AU} 
  \\ 
  & \multicolumn{4}{c}{$0 \leq z \leq 6.6$ AU} &
  \multicolumn{2}{c}{$0 \leq z \leq 13.2$ AU} 
  \\ 
AMR max. resolution & \multicolumn{6}{c}{$4\times 10^8$ cm ($2.7\times 10^{-5}$ AU)} \\
Time covered & \multicolumn{6}{c}{0--18 hours} \\ \hline
\end{tabular}
\end{center}
\label{default}
\end{table*}%

\begin{figure}
\begin{center}
  \includegraphics[angle=0,width=0.45\textwidth]{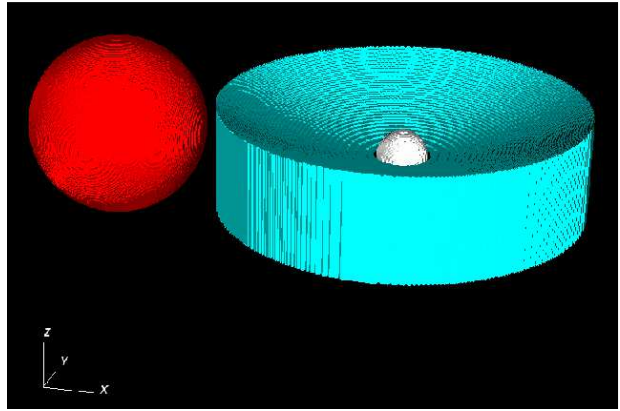}
\end{center}
\caption{A rendition of  the U~Sco binary system model initial conditions.
The white sphere represents the
pre-outburst, initial radius of the Sedov-type blast on the white dwarf.
}
\label{f:preblast}
\end{figure}

For the system parameters, we adopt the values of
\citet{Thoroughgood.etal:01}; these are listed in Table~\ref{t:params},
together with the parameters adopted for the blast models investigated
here. We included a diffuse, puffed-up disk-like distribution
around the system which is analogous to the gas density enhancement found
in the numerical model of the RS~Oph system immediately prior to outburst
by \citet[][see also \citealt{Mastrodemos.Morris:99}]{Walder.etal:08}
and to the equatorial gas distribution adopted by
\citet{Orlando.etal:09}.
%
% the particle density distribution of the unperturbed CSM is given by}
% 
% \begin{equation}
% n = n_{\rm w} + n_{\rm eq}~\exp\left[-\left(\frac{x-x_0}{16 L}\right)^2 -
% \left(\frac{y-y_0}{16 L}\right)^2 - \left(\frac{z-z_0}{L}\right)^2\right]~,
% \end{equation}
% 
% \noindent
% \salvo{where $n_{\rm w}$ is the density of the secondary star wind,
% $n_{\rm eq}$ is the density enhancement at the equatorial plane, $(x_0,
% y_0, z_0)$ are the coordinates of the secondary star, and $L=5\times
% 10^{6}$~km is a characteristic length scale. 
%
We also included a flared, uniform density gas accretion disk 
around the white dwarf, adopting the shape from \citet{Hachisu.etal:00}.
We included the secondary star as an impenetrable body.  A close-up
of this initial pre-blast configuration is illustrated in
Figure~\ref{f:preblast}.

As an initial condition, we assumed a spherical Sedov-type blast wave
\citep{Sedov:59} originating from the thermonuclear explosion centered
on the white dwarf, with radius $r_{\rm 0} = 4\times 10^5$~km
and total energy $E_{\rm 0}$. This energy was partitioned such
that 1/4 was contained in thermal energy, and 3/4 in kinetic energy.
Solar abundances of \citet{Grevesse.Sauval:98} (GS) were assumed for the
disk, wind and equatorial ambient gas, while the ejecta metallicity was assumed
enhanced by a factor of ten. This latter choice was guided by observations of He-rich ejecta in previous outburst, as noted in Sect.~\ref{s:intro}, as well as 
the \citet{Drake.etal:09} high-resolution X-ray spectroscopic study
of the 2006 RS~Oph blast that found evidence for metal-rich ejecta.
Since radiative losses were not important for the early blast evolution,
the choice of abundances is only relevant for computation of the emitted
X-ray intensity of the blast.

Six models were computed to explore the effects of a different
accretion disk and circumstellar gas density, explosion energy and total ejecta
mass on the outcome of the blast. Direct estimates of the 
ejecta mass from observations of the 1979 and 1999 outbursts
\citep{Williams.etal:81,Anupama.Dewangan:00,Evans.etal:01}
find $M_{ej}\sim10^{-7}M_{\odot}$. Expansion velocities of $\sim
5000$~km~s$^{-1}$ from optical spectroscopy correspond to a kinetic
energy of $2.5\times 10^{43}$~erg, which can be taken as a lower
limit to the explosion energy.  This is consistent with estimates of
the integrated optical output from earlier outbursts of $10^{44}$~erg
\citep[e.g.][]{Webbink.etal:87}. We therefore adopted fiducial values of
$E_0=10^{44}$~erg and $M_{ej}=10^{-7}M_{\odot}$ and explored the effects
of lower and higher values of each by an order of magnitude. For the
accretion disk we assumed a uniform density of $n_{d}=10^{15}$~cm$^{-3}$
and investigated a further model with $n_{d}=10^{13}$~cm$^{-3}$.
The wind and equatorial density enhancement are included for 
completeness.  While this circumstellar gas is not expected to have a profound influence on the dynamics and evolution of the blast, it does influence the shock structure and 
the predicted X-ray luminosity.   We probed the effects of increasing and decreasing 
the circumstellar gas density by a factor of 10.

The hydrodynamic equations were solved in
one quadrant of the 3-D spatial domain and the coordinate system was oriented
in such a way that both the white dwarf and the donor star lie on the
$x$ axis. The donor star was at the origin of the coordinate system,
$(x_0,y_0,z_0) = (0,0,0)$, and the white dwarf was located at $x = 0.03$ AU, 
i.e. the system orbital separation.  The extents of spatial domains employed for the different models are listed in Table~\ref{t:params} together with other relevant parameters, and depended on the circumstellar gas density; two models that investigated lower densities for which expansion rates were greater required larger domains.
%and the computational domain extended 13.2 AU
%in the $x$ direction, and 6.6 AU in both the $y$ and $z$ directions;

The explosion was followed for 18 hours in order to explore the early
phase X-ray emission and capture the details of the effects of the
explosion environment on the blast evolution. A major challenge
in modelling the explosion of U~Sco is the very small scale of the
system compared with the size of the rapidly expanding blast wave.
The separation of the two stars is only 0.03 AU---a distance traversed
by a blast wave moving at 5000~km~s$^{-1}$ in less than 20 minutes---and
the secondary star subtends a sufficiently large solid angle that it
has a significant effect on the blast wave evolution. To capture this
range of scales, the models explored here employed 16 nested levels
of adaptive mesh refinement, with resolution increasing twice at each
refinement level. This grid configuration yields an effective resolution
of $\approx 4\times 10^3$ km at the finest level, corresponding to
$\approx 100$ grid points per initial radius of the blast.

%Since we do not expect these density values to be comparable with
%the disk density, we did not explore the effects of these parameters
%on our simulations.

Given the parameters in Table~\ref{t:params}, for five of our six models 
$t_{\rm tr} > 3$~days from Eq.~\ref{trans_time}, and for the sixth (E44M-7D15HW) $t_{\rm tr}=2.3$~days.  These times are 
much larger than the time covered by our simulations.  Our
modeled remnants therefore never enter the radiative phase.

%helium-enriched model poses a serious problem. Truran et al. 
%(1988) discussed the composition dependence of thermonuclear runaway mod- 
%els for recurrent novae of U Sco-type. They showed that for Mwd = 1.38 M ? ,

%Msg = 1.5 ? 10?8 M ? /yr and L = 0.1 L ? optically bright outbursts are obtained 
%only for matter with He/H < 1 by mass.

\section{Results and Discussion}
\label{s:discuss}

\subsection{Temperature and density structure}
\label{temp_dens_sec}

The U~Sco models all evolved quite differently to those presented for RS~Oph by \citet{Orlando.etal:09}.  As noted in \S1, this difference is expected because U~Sco lacks the dense wind of RS~Oph for propagation of an X-ray emitting blast wave.

\begin{figure*}
\begin{center}
  \includegraphics[angle=0,width=1.0\textwidth]{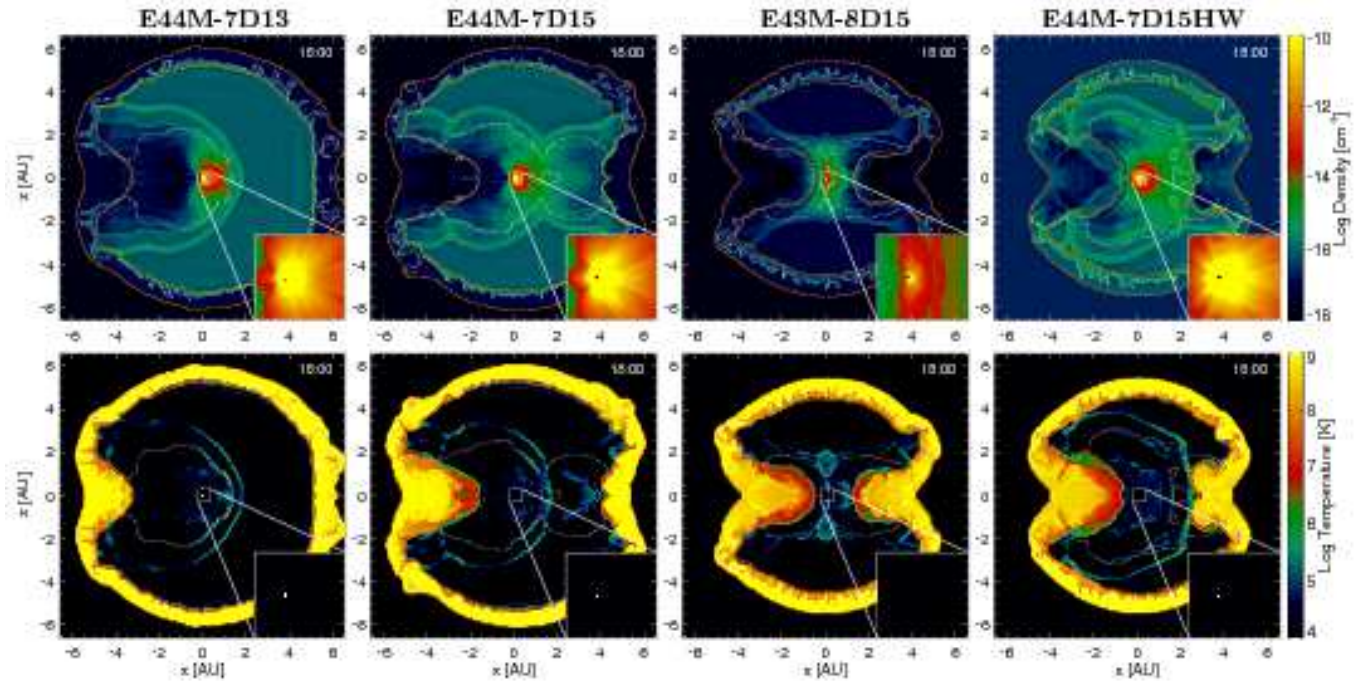}
\end{center}
\caption{
Colour-coded cross-section images of the gas density distribution (top)
and temperature (bottom) 18 hours after the U~Sco blast, corresponding
to the end of the hydrodynamic simulations.  The secondary star is
at the origin, and the white dwarf to the right.  The different 
panels illustrate models with accretion disk density $n_{\rm d} =
10^{13}$~cm$^{-3}$ and an explosion energy of  $E_0 = 10^{44}$ erg
(E44M-7D13; Left); $n_{\rm d} = 10^{15}$~cm$^{-3}$
and $E_0 = 10^{44}$ erg (E44M-7D15; left-center); 
$n_{\rm d} = 10^{15}$~cm$^{-3}$ and $E_0 = 10^{43}$ erg (E43M-8D15; right-center).  Far right panels show a similar model to the left-center panels, except with CSM 
gas densities increased by a factor of ten (E44M-7D15HW).  Inset
panels show the blast structure closer to the system origin.  The
secondary star is shown and is responsible for the strong ``shadowing"
of the blast wave. The accretion disk is completely destroyed by the
blast in all simulations. The white dotted contour encloses the ejecta. The red
solid contour denotes the regions with plasma temperature $T> 1\times
10^6$~K.
}
\label{f:blast}
\end{figure*}

The 18 hour post-blast gas density and temperature distributions in
the $x$-$z$ plane bisecting the system (the plane of the orbital axis;
the accretion disk is edge on) are illustrated in Figure~\ref{f:blast}.
All the models exhibit conspicuous departure from spherical
symmetry. The main feature of the gas density distribution is a large
cavity to the left of the figures (negative $x$ direction) which is a
blast wave ``shadow'' cast by the secondary star. This result reinforces
the need to perform simulations with sufficient spatial resolution to
capture the details of the pre-blast system configuration.

In all models the accretion disk is completely destroyed by
the blast.  This is not surprising considering the gravitational binding energy of the disk in our models is 3-4 orders of magnitude lower than the explosion energy:  disk destruction is assured unless the disk gas density is substantially higher than $10^{15}$~cm$^{-3}$.
\citet[][see also \citealt{Munari.etal:10}]{Worters.etal:10} report a renewed optical flickering with amplitude ``0.2~mag
over the course of an hour'' on 2010 February 5.  They interpret this 
re-establishment of accretion, implying a limit of 7 days on
the build-up of an angular momentum shedding accretion disk.  In
our simulations, the disk volume is $V_{\rm disk}\approx 2\times
10^{34}$~cm$^{3}$ and its mass ranges between $2.2\times 10^{-10}
M_{\odot}$ (model E44M-7D13) and $2.2\times 10^{-8} M_{\odot}$ (all other models).  
%models E44M-7D15, E43M-8D15 and E44M-7D15HW). 
The build up of the disk in 7 days 
would imply a rate of mass loss from the companion star ranging between
$10^{-8}$ and $10^{-6}M_{\odot}$~yr$^{-1}$, consistent with the estimate of \citet{Hachisu.etal:00}.  We note, however, that any renewed disk build-up would need to overcome the dynamic pressure of the radiatively-driven outflow from the evolving central object.

\begin{figure}
\begin{center}
  \includegraphics[angle=0,width=0.45\textwidth]{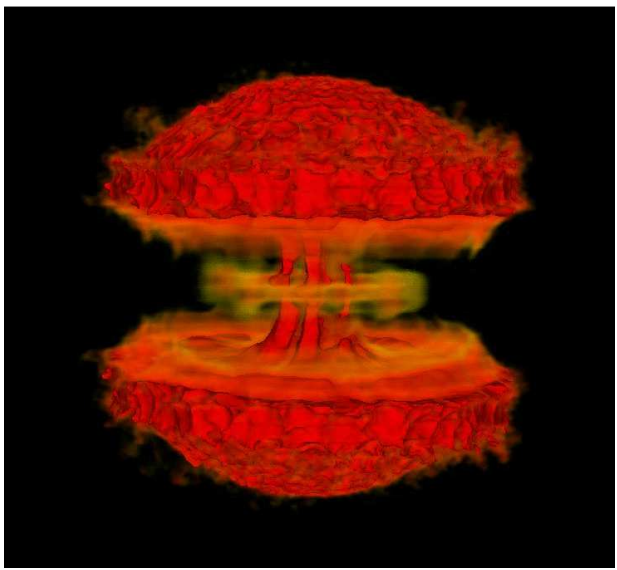}
\end{center}
\caption{
Three-dimensional rendering of the ejecta distribution 18 hours
after the blast for the model E44M-7D15. The plane of the orbit of
the central binary system lies on the $(x,y)$ plane.
}
\label{f:ejecta}
\end{figure}

Model E44M-7D13, with lower accretion disk gas density, shows only
a very small effect of the disk  in a slight inhibition of
the blast wave progress in the positive $x$ direction. In contrast,
the higher disk gas density in all 
the other models has quite a profound effect on the blast evolution.  
Both the blast and ejecta are strongly collimated in polar directions,
the collimation being more prominent for lower explosion energy
(E43M-8D15) as might be expected. 
Figure~\ref{f:ejecta} shows, as an example, the collimation
of ejecta 18
hours after the outburst for the model E44M-7D15.
\citet{Kato.Hachisu:03}
interpreted ``horned'' optical line profiles observed in previous  U~Sco
outburst \citep[e.g.][]{Munari.etal:99}
as evidence for ``jet-shaped'' outflows. Similar blast
collimation was both predicted by hydrodynamic simulations of the early
RS~Oph blast by \citet{Walder.etal:08} and \citet{Orlando.etal:09}, and
deduced through X-ray spectroscopy obtained 13.6 days after the outburst
\citep{Drake.etal:09}. The simulations indicated the collimation was
caused by an equatorial disk-like gas distribution around the white dwarf.
High spatial resolution observations of the blast at various times
also found optical, infrared, radio and X-ray collimation signatures
\citep{O'Brien.etal:06,Bode.etal:07,Chesneau.etal:07,Luna.etal:09}.

Evidence for a more strongly collimated synchrotron radio jet was also
found by \citet{Rupen.etal:08} and \citet{Sokoloski.etal:08}. The
latter authors concluded that such structure is too highly collimated to
be explained by the influence of circumbinary material on the explosion.
The calculations presented here and by \citet{Orlando.etal:09} bring
such conclusions into question: the nature of the explosion environment
is crucial for the morphology of the evolving blast and ejecta.
While the models presented here do not produce obvious narrow jets
of material, we note that radio synchrotron emission also depends
strongly on the magnetic field strength, gradient and orientation
\citep[e.g.,][]{Orlando.etal:07} and need not be co-spatial with obvious density enhancements in the ejecta.
%It seems plausible that a jet-like
%radio appearance might be produced by, for example, a combination
%of poleward hydrodynamic collimation combined with a magnetic field
%stronger along the axial direction. 

The outer regions of the blast are characterised by a layer of
shocked circumstellar medium (CSM) with temperature $\approx 10^{9}$~K
and a thin expanding surface of shocked ejecta with temperature in the range 
$[10^7 - 10^8]$~K (see Figure~\ref{f:blast}).   The 
CSM gives rise to a reflected shock that heats the ejecta, and is strongest for model E44M-7D15HW in which the CSM density is highest.  The
secondary star also gives rise to a prominent ``bow shock'', though
in this structure the gas is shocked to a temperature of only a few
$10^{5}$~K and does not give rise to any conspicuous observable X-ray
emission. It is worth noting that Fig.~\ref{f:blast} reports the proton
temperature set by collisionless heating at the shock front. Given
the short timescale considered here, protons and electrons are
likely not thermalized at the forward shock, so that the ratio
of the electron to proton temperature at the blast wave is expected
to be $(T_{\rm e}/T_{\rm p}) < 1$. \citet{Ghavamian.etal:07} derived a
physical model for the heating of electrons and protons in strong
shocks that predicts a relationship $(T_{\rm e}/T_{\rm p}) \propto
v_{\rm S}^{-2}$, where $v_{\rm S}$ is the shock velocity.
%, that is fully consistent with the observations. 
Considering the velocity of the
forward shock in our simulations is of the order of 10000~km~s$^{-1}$,
from Fig.~2 of \citet{Ghavamian.etal:07} we derive $(T_{\rm e}/T_{\rm p})
< 0.1$.  Thus, we expect that the electron temperature of the shocked
CSM is $\approx 10^8$~K.  X-rays are instead naturally
dominated by the higher density parts of the shocked ejecta which are
expected to equilibrate quickly. In the following, we therefore assumed
electron-proton temperature equilibration in shocked ejecta.

\begin{figure*}
\begin{center}
  \includegraphics[angle=0,width=1.\textwidth]{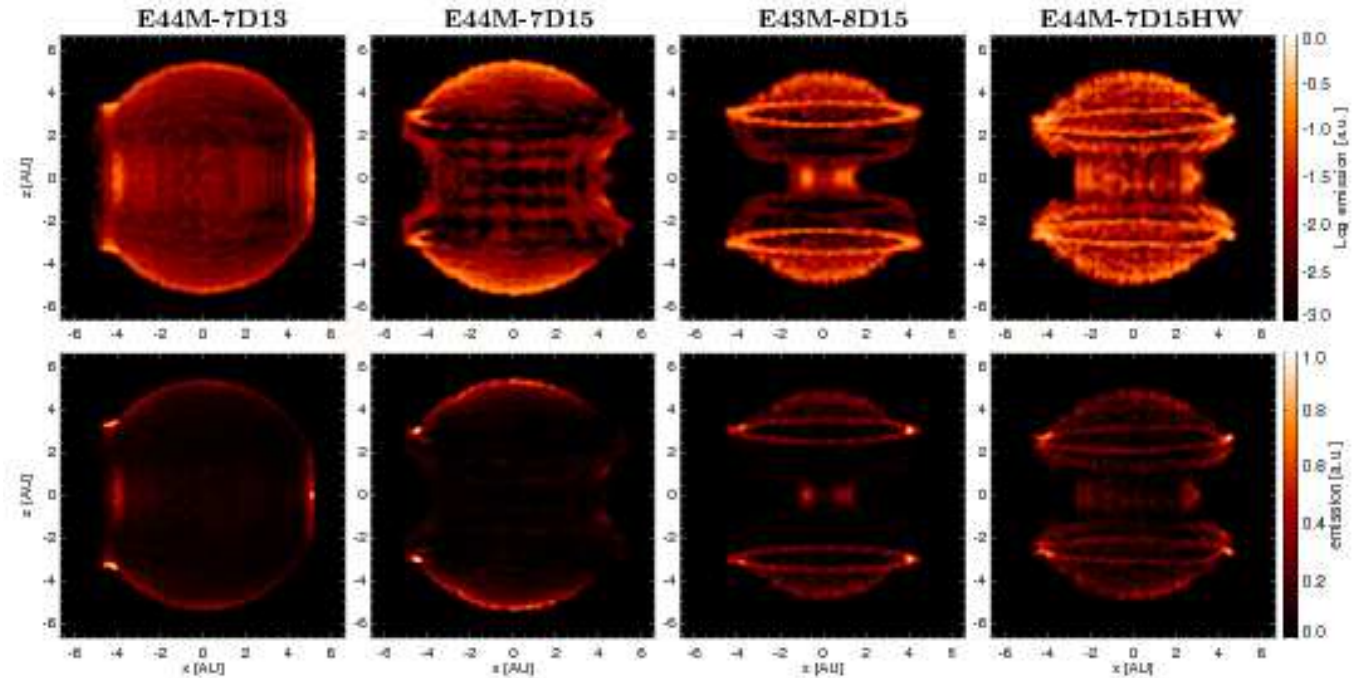}
\end{center}
\caption{
X-ray images (normalized to the maximum of each panel) projected
along the line of sight of the blast after 18 hours of evolution,
corresponding to the density and temperature distributions illustrated
in Figure~\ref{f:blast}. 
%The ordering of the images is the same as that in Figure~\ref{f:blast}.
%and is, from left to right, E44M-7D13, E44M-7D15, E43M-8D15 and E44M-7D15HW. 
Upper panels are rendered with a logarithmic surface
brightness scale and lower panels with a linear scale. The plane of
the orbit is inclined by 7 degrees to the line-of-sight. Both the secondary 
subgiant and the white dwarf lie on the x axis.  Most of the X-ray emission
originates from shocked ejecta.
}
\label{f:xray}
\end{figure*}

\subsection{X-ray luminosity and morphology}

X-ray emission from
the blast was synthesised from the model results using the methodology described by \citet{Orlando.etal:09}. The synthesis includes
thermal broadening of emission lines, the Doppler shift of lines
due to the component of plasma velocity along the line-of-sight,
and photoelectric absorption by the interstellar medium (ISM),
CSM, and ejecta. The absorption by the ISM is calculated assuming
a column density $N_{\rm H} = 1.4\times10^{21}$ cm$^{-2}$ (e.g.,
\citealt{Kahabka.etal:99}, and is consistent with a distance to U~Sco of $12$~kpc,
e.g. \citealt{Schaefer:10}); the local absorption is calculated as due to
shocked CSM (assuming GS abundances) and ejecta (GS abundances $\times
10$).  The X-ray emission from the blast at 18 hours is illustrated for four representative models in
Figure~\ref{f:xray}.  The absorption by the ejecta contributes to a small
difference between the predicted X-ray luminosities of the different
models. Most of the emission arises from the high-temperature shocked
ejecta forming the outer ``shell'' of the blast wave where we
assumed electron-proton temperature equilibration.  Model E44M-7D13
exhibits significant equatorial brightening that arises from a ring
of higher density ejecta that was partially impeded by the accretion
disk. This structure is not present in models E44M-7D15, 
E43M-8D15 and E44M-7D15HW, with a higher density disk. Instead, these latter models
are dominated by bipolar cusps that appear either limb-brightened
(E44M-7D15), or forming ring-like structures (E43M-8D15 and E44M-7D15HW), in the projected image, accompanied by  limited equatorial emission.

The X-ray luminosities of the simulations in the
[0.6--12.4]~keV band are listed in Table~\ref{t:params} and vary between 
$L_{X}=3\times 10^{31}$~erg~s$^{-1}$ (E43M-7D15) and $7\times 10^{33}$~erg~s$^{-1}$ (E45M-6D15LW).  The X-ray luminosity is larger for higher ejecta mass, explosion energy, and circumstellar gas density.  
%$L_{X}=1\times 10^{33}$~erg~s$^{-1}$
%(E44M-7D13), $5\times 10^{32}$~erg~s$^{-1}$ (E44M-7D15), $3\times
%10^{31}$~erg~s$^{-1}$ (E43M-8D15), and $4\times 10^{33}$~erg~s$^{-1}$ 
%(E44M-7D15HW).  
The X-ray flux upper limit of $F_X <
1.1\times 10^{-13}$~erg~s$^{-1}$cm$^{-2}$
obtained at a similar epoch to our simulations by \citet{Schlegel.etal:10}
corresponds to an X-ray luminosity of $L_{X}=2\times 10^{33}$~erg~s$^{-1}$
for a distance of 12~kpc.  The fact that the models with $M_{\rm ej} =
10^{-7}M_{\odot}$  predict X-ray luminosities
close to the {\it Swift} upper limit (except run E44M-7D15LW whose luminosity is much lower), while the model with $M_{\rm ej}=10^{-6}M_{\odot}$ and lower circumstellar density (E45M-6D15LW) exceeds  it, 
 suggests that the mass of ejecta
in the 2010 outburst was not significantly larger than $10^{-7}M_{\odot}$.   
This is consistent with estimates of previous outbursts in 1979 and 1999
\citep{Williams.etal:81,Anupama.Dewangan:00,Evans.etal:01}.  While we cannot rule out with certainty the circumstellar density being even lower than assumed in the latter model with $M_{\rm ej}=10^{-6}M_{\odot}$, which could reduce the predicted X-ray luminosity and reconcile it with the {\it Swift} upper limit, such a large mass requires the larger assumed explosion energy of $10^{45}$~erg to take the ejecta out of the gravitational well of the system.  This exceeds by an order of magnitude the  total explosion energy of $10^{44}$~erg estimated for previous outbursts by \citep{Webbink.etal:87}.

The model with a higher wind and equatorial gas density (E44M-7D15HW) has a stronger X-ray contribution from ejecta heated in the reflected shock that {\em exceeds} the observed X-ray upper limit.  This is consistent with our assumption of a rather low circumstellar gas density, commensurate with the evolutionary state of the U~Sco component stars, and 
does not alter the conclusion that the ejected mass was probably not larger than 
$10^{-7}M_{\odot}$.  

%As a
%consequence, the white dwarf is increasing in mass if its growth rate
%is larger than $10^{-8}~M_{\odot}$~yr$^{-1}$ (the period of the nova
%is $\approx 10$ years). Since the rate of mass loss from the donor star
%derived in Sect.~\ref{temp_dens_sec} is $> 10^{-8}~M_{\odot}$~yr$^{-1}$,
%we argue that the white dwarf can efficiently grow in mass towards the
%Chandrasekhar limit and ultimately explode as a SN Ia.}

%\subsection{Future Evolution}
%
%The results obtained on the early blast provide some predictive power
%for future high spatial resolution observations. We conclude that the
%presence of a substantial accretion disk {\em will} result in bipolar
%structure. 

\acknowledgments

JJD was funded by NASA contract NAS8-39073 to the {\it Chandra X-ray
Center}. 
%during the course of this research and thanks the CXC
%director, Harvey Tananbaum, and the CXC science team for advice and
%support. 
%The software used in this work was in part 
FLASH was developed by
the DOE-supported ASC/Alliance Center for Astrophysical Thermonuclear
Flashes, University of Chicago.  Simulations were executed at
the HPC SCAN facility of INAF-OAPA.   

\end{document}